\documentclass[twocolumn]{aastex631}

\usepackage{threeparttable}

\submitjournal{AJ}

\shorttitle{Peculiar orbital characteristics of 469219 Kamo`oalewa}
\shortauthors{Hu et al.}

\begin{document}

\title{Peculiar orbital characteristics of Earth quasi-satellite 469219 Kamo`oalewa: implications for the Yarkovsky detection and orbital uncertainty propagation}

\correspondingauthor{Shoucun Hu}
\email{hushoucun@pmo.ac.cn, jijh@pmo.ac.cn}

\author{Shoucun Hu}
\affil{CAS Key Laboratory of Planetary Sciences, Purple Mountain Observatory, Chinese Academy of Sciences, Nanjing 210023, China\\}
\affil{CAS Center for Excellence in Comparative Planetology, Hefei 230026, China\\}
\affil{School of Astronomy and Space Science, University of Science and Technology of China, Hefei 230026, China\\}

\author{Bin Li}
\affil{CAS Key Laboratory of Planetary Sciences, Purple Mountain Observatory, Chinese Academy of Sciences, Nanjing 210023, China\\}
\affil{CAS Center for Excellence in Comparative Planetology, Hefei 230026, China\\}
\affil{School of Astronomy and Space Science, University of Science and Technology of China, Hefei 230026, China\\}

\author{Haoxuan Jiang}
\affil{CAS Key Laboratory of Planetary Sciences, Purple Mountain Observatory, Chinese Academy of Sciences, Nanjing 210023, China\\}

\author{Gang Bao}
\affil{CAS Key Laboratory of Planetary Sciences, Purple Mountain Observatory, Chinese Academy of Sciences, Nanjing 210023, China\\}

\author{Jianghui Ji}
\affil{CAS Key Laboratory of Planetary Sciences, Purple Mountain Observatory, Chinese Academy of Sciences, Nanjing 210023, China\\}
\affil{CAS Center for Excellence in Comparative Planetology, Hefei 230026, China\\}
\affil{School of Astronomy and Space Science, University of Science and Technology of China, Hefei 230026, China\\}

\begin{abstract}

469219 Kamo`oalewa is selected as one of the primary targets of Tianwen-2 mission, which is currently believed to be the most stable quasi-satellite of Earth. Here we derive a weak detection of the Yarkovsky effect for Kamo`oalewa, giving $A_2 = -1.075\pm0.447\times 10^{-13} \rm{au/d}^2$, with the available ground-based optical observations from Minor Planet Center and a relatively conservative weighting scheme. Due to the quasi-satellite resonance with Earth, we show that the detection of Yarkovsky effect by orbital fitting with astrometric observations becomes difficult as its orbital drift shows a slow oscillatory growth resulting from the Yarkovsky effect. In addition, we extensively explore the characteristics of orbital uncertainty propagation and find that the positional uncertainty mainly arises from the geocentric radial direction in 2010-2020, and then concentrates in the heliocentric transverse direction in 2020-2030. Furthermore, the heliocentric transverse uncertainty is clearly monthly dependent, which can arrive at a minimum around January and a maximum around July as the orbit moves towards the leading and trailing edges, respectively, in 2025-2027. Finally, we investigate a long-term uncertainty propagation in the quasi-satellite regime, implying that the quasi-satellite resonance with Earth may play a crucial role in constraining the increase of uncertainty over time. Such interesting feature further implies that the orbital precision of Kamo`oalewa is relatively stable at its quasi-satellite phase, which may also be true for other quasi-satellites of Earth.

\end{abstract}

\keywords{minor planets, asteroids: general --- astrometry --- ephemerides --- methods: numerical}

\section{Introduction} \label{sec:intro}
Near-Earth asteroids are defined as those with perihelion distances less than 1.3 au. Among them, there is a special classification called Earth co-orbital asteroids that are captured in a 1:1 mean-motion resonance with Earth \citep{namouni1999secular}. Such asteroids have three typical orbital configurations, i.e., tadpole orbit that moves about the Lagrangian points L4 and L5, {horseshoe orbit librating with a large amplitude (as large as to around 180$^\circ$) in longitude in the frame co-rotating with the Earth, and quasi-satellite orbit librating around 0$^\circ$ (typically with small amplitudes)} \citep{morais2002population, connors2002discovery}. Under the influence of complex gravitational environment, the orbits may periodically switch between horseshoe and quasi-satellite configurations \citep{sidorenko2014quasi, qi2022co}.

469219 Kamo`oalewa (provisional designation is 2016 HO3) is a fast-rotating near-Earth asteroid discovered by the Pan-STARRS1 survey telescope at the Haleakala Observatory in Hawaii on 27 April 2016 \citep{tholen2016potpourri, de2016asteroid}. The asteroid has a quasi-satellite orbit with respect to Earth, with a semi-major axis of 1.001 au, an eccentricity of 0.103 and an orbital inclination of 7.79$^\circ$. Spectral observations indicate that Kamo`oalewa may comprise lunar material, and the albedos ranging from 0.10 to 0.16 give an estimate of the effective diameter of 58-46 m \citep{sharkey2021lunar, winter2022possibility}. The amplitude of light curves suggest that the shape of this object is probably elongated, with a length-to-width ratio of less than 0.48 \citep{li2021shape}. The spin period of Kamo`oalewa is 28.3$_{-1.3}^{+1.8}$ minutes \citep{sharkey2021lunar}, which is lower than the critical value of $\sim$2.2 hours \citep{pravec2000fast}, implying that internal cohesion is required to maintain its internal structure \citep{li2021shape, hu2021critical}.

The orbit of Kamo`oalewa exhibits both retrograde and prograde motions in the geocentric inertial frame, while it turns entirely retrograde in the geocentric co-rotating frame \citep{pousse2017co} as shown Fig.\ref{fig:fig00_HO3_orbit_when_obs}. Orbital propagation shows that its quasi-satellite state remains very stable, with a geocentric distance oscillating in the range 0.1-0.3 au (outside the Earth's Hill radius 0.01 au). {Long-term  simulations show that its orbit periodically switches between horseshoe and quasi-satellite configurations, and its current orbit state began about 100 years ago and will transition back to the horseshoe orbit in roughly 300 years} \citep{de2016asteroid, sharkey2021lunar}. Kamo`oalewa is a good target for low-cost in-situ study \citep{heiligers2019trajectory, venigalla2019near} due to its peculiar orbit relative to the Earth. In addition, China National Space Administration (CNSA) will launch a sample-return mission to this asteroid {around} 2025 \citep{chi2018power, huang2020small, zhang2021china, zhang2021developing}, which will be the second mission of China to closely rendezvous a near-Earth asteroid \citep{huang2013ginger}. Numerical simulations with ground-based and on-board radiometric tracking data were performed to estimate the GM value, which is a major challenge due to the tiny mass of the asteroid \citep{jin2020simulated,yan2022simulation}.

{The} Yarkovsky effect is a radiation recoil force acting on rotating asteroids caused by anisotropic thermal emission of absorbed sunlight \citep{vokrouhlicky2000yarkovsky}, which plays a {fundamental} role as a non-gravitational force in the orbital evolution of near-Earth asteroids, especially for smaller objects \citep{ bottke2002effect, morbidelli2003yarkovsky, bottke2006yarkovsky}. The transverse component of {Yarkovsky acceleration gives rise to a typically slow} variation in the semi-major axis, as well as a long-term drift in the mean anomaly \citep{farinella1998meteorite}. {Long term dynamical simulations that included the Yarkovsky effect suggest that the mechanism can cause an earlier removal of the asteroid from current co-orbital configuration than that of a gravity-only model does \citep{fenucci2021role}.}

The Yarkovsky acceleration can vary as a function of mass, spin state and surface thermal properties \citep{vokrouhlicky1999complete}. Direct calculation of the force by thermophysical modeling is usually unrealistic due to the unknown parameters for a vast majority of near-Earth asteroids. The effect for a 1-km-sized object generally results in a subtle drift of $\sim{10^{ - 3}}{\rm{au}}/{\rm{My}}$ in the semi-major axis, with a related magnitude of acceleration close to the order of gravitational perturbation from the main belt \citep{nugent2012detection}. {If accurate} radar observations are available or optical observations cover {long} arc, this effect can be detected from orbital fitting \citep{nugent2012detection, farnocchia2013near, greenberg2017asteroid, deo2017yarkovsky, del2018detecting, liu2022yarkovsky}. \cite{greenberg2020yarkovsky} presented the detection of Yarkovsky effect for 247 near-Earth asteroids with optical and radar astrometry, and high precision observations will reveal substantial {detections} \citep{desmars2015detection, chesley2015direct, dziadura2022investigating}. If the {shape}, diameter and spin state of the asteroid are measured from Mid-IR observations, the bulk density {and even the thermophysical properties such as thermal inertia may be further unveiled by the Yarkovsky drift \citep{chesley2003direct, rozitis2011directional, rozitis2013thermophysical, rozitis2014physical, chesley2014orbit, hanuvs20183200, jiang2019Revisiting, fenucci2021low}}.

{Recently, \cite{liu2022yarkovsky} presented the Yarkovsky detection for Kamo`oalewa using a modified open source software. Here we aim to perform a comprehensive study of the Yarkovsky detection and understand orbital uncertainty propagation characteristics for Kamo`oalewa. In addition, we employ our developed orbital determination package that adopts a slightly more conservative consideration of weighting scheme to conduct orbital fitting, resulting in a weak detection of Yarkovsky drift of the asteroid. Based on a series of orbital determination simulations, we further show that Kamo`oalewa's peculiar quasi-satellite orbit makes it difficult to detect the Yarkovsky effect with ground-based optical observations. We then study the characteristics of the orbital uncertainty in the next few years to provide information for Tianwen-2 mission. Furthermore, we find that the uncertainty propagation of Kamo`oalewa in the quasi-satellite regime can be constrained by the quasi-satellite resonance with Earth, which is an interesting feature for such a stable quasi-satellite.}

The paper is structured as follows. Section \ref{Sec2} describes the adopted observations for orbital fitting, along with the dynamical model and detection method of Yarkovsky effect. In Section \ref{Sec3}, we show the orbital behavior of Kamo`oalewa involved in Yarkovsky effect, and derive a weak detection of Yarkovsky effect from the fitting. In Section \ref{Sec4}, we explore the characteristics of orbital uncertainty with given orbital solutions, and carry out a secular investigation in the quasi-satellite regime. Finally, we summarize the main results in Section \ref{Sec5}.

\section{Observations and Models}
\label{Sec2}
\subsection{Ground-based optical astrometry}
\begin{figure*}
    \centering
    \includegraphics[width=\textwidth]{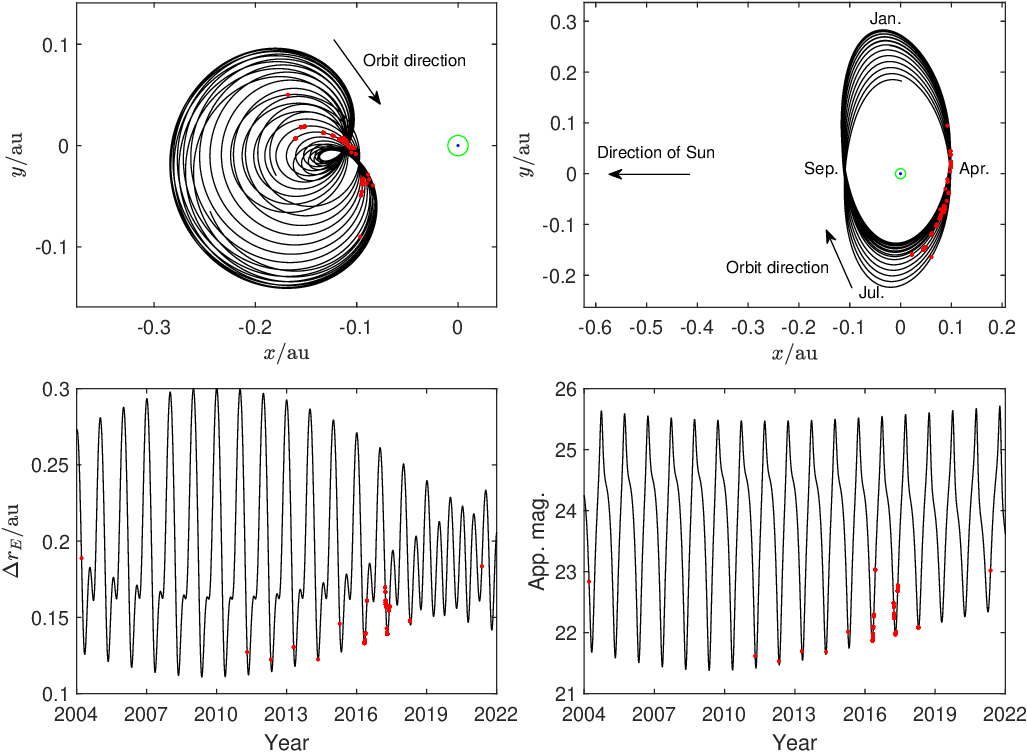}
    \caption{The upper panels show the geocentric orbit of Kamo`oalewa in the mean ecliptic reference frame of J2000.0 and the Sun-Earth rotating frame, respectively (both are projected from the direction of the north ecliptic pole). The lower panels show the variations in the geocentric distance and apparent magnitude, respectively. The time span ranges from 2004 to 2022. The red dots mark up the observation epochs, while the green circles represent the Earth's Hill sphere.}
    \label{fig:fig00_HO3_orbit_when_obs}
\end{figure*}

{Considering the small size and relatively large geocentric distance ($\Delta r_E$) of Kamo`oalewa, ground-based radar observations are unavailable}. However, its stable orbit with respect to Earth allows it to have a good observational window around April once a year. As of 2022 November 1, 310 measurements of {right ascension (RA) and declination (DEC)} spanning from 2004 March 17 to 2021 May 13 were released from 10 different observatories at MPC, among which the observations by the station T12 {(University of Hawaii 2.24-m telescope, Maunakea)} contributes a large portion of 70\%. Fig.\ref{fig:fig00_HO3_orbit_when_obs} shows the orbit in the geocentric ecliptic frame, geocentric rotating frame, time evolution of $\Delta r_E$ and apparent magnitude from 2004 to 2022, where the red dots correspond to the observation epochs (see also \cite{de2016asteroid}).

\subsection{Yarkovsky modeling}
The Yarkovsky effect can be an important source of uncertainty in the orbit propagation of a near-Earth asteroid \citep{chesley2014orbit}. Generally, only the transverse component of the force needs to be considered in the modeling. The following simple expression is adopted to quantify the force \citep{del2018detecting}
\begin{equation}
    \label{eq:Yarkovsky_force}
    {{\bf{F}}_{\rm{Yar}}} = {A_2}{\left( {\frac{{{r_0}}}{r}} \right)^d}{\bf{\hat t}},\;\;\;0.5 \le d \le 3.5
\end{equation}
where $r_0=1\;\rm{au}$ and $d$ is a parameter related to the thermophysical properties. Here $d = 2$ is assumed, though its specific value has little effect on the orbit since the heliocentric distance $r$ of Kamo`oalewa is always around 1 au. ${{\bf{\hat t}}}$ is a unit vector pointing to the transverse direction of the orbit and {$A_2$ is a free parameter, reflecting} the strength of the Yarkovsky acceleration. The mean semi-major axis drift rate can then be expressed as {\citep{farnocchia2013near}}
\begin{equation}
    \label{eq:dadt}
    \langle {\rm{d}}a/{\rm{d}}t \rangle  = \frac{{2{A_2}\left( {1 - {e^2}} \right)}}{n}{\left( {\frac{{{r_0}}}{p}} \right)^2}
\end{equation}
{where $a$, $e$ and $n$ denote the semi-major axis, the eccentricity, and the mean motion, respectively, and $p$ is the semi-latus rectum $a(1-e^2)$.}

\subsection{Dynamical model}
\label{sub:dynamical model}
To detect the Yarkovsky effect, a high fidelity dynamical force model is required to compute the orbit. Here the Newtonian accelerations of the Sun (as well as the relativistic effect), the eight major planets, the Moon and the 16 massive main-belt asteroids are considered \citep{farnocchia2013near}, using the JPL planetary ephemeris DE440 to calculate the positions of the planets during the orbital computation \citep{park2021jpl}. Since the orbit is limited to around 1 au, the uncertainty caused by other massive asteroids is {negligible} at the current accuracy. In addition, the oblateness perturbation of the Sun and the Earth, as well as the solar radiation pressure, can be also safely ignored at the current level of observation accuracy \citep{liu2022yarkovsky}.

\subsection{Method of Yarkovsky effect detection}
\label{sub:method of yarkovsky effect detection}
{In this work}, we perform the orbital fitting with a 7-dimensional differential correction in both the orbital elements and $A_2$ simultaneously from the dataset \citep{farnocchia2013near}. Here we have developed a software called SBORD (Small-Body ORbit Determination package) to perform orbital determination with ground-based optical astrometry. A weighted least-squares algorithm is adopted to fit the astrometric measurements \citep{milani2010theory}. The latest debiasing technique by \cite{eggl2020star} and the weighting scheme by \cite{verevs2017statistical} are implemented to process the data. {The outlier rejection scheme adopted is the same as \cite{greenberg2020yarkovsky}}. {The Runge-Kutta-Fehlberg (RKF78) method is used to propagate the orbit of the asteroid \citep{fehlberg1969classical}.}

To examine the reliability of SBORD, we perform an orbital fit for the near-Earth asteroid (499998) 2011 PT with the observations from 2011 to 2017 given by MPC. {Next, we measure the Yarkovsky drift of $A_2$ = (-2.133 $\pm$ 0.296)$\times10^{-13}$ au/d$^2$ for 2011 PT (the epoch is 2023 Feb. 25 TDB), which agrees well with the JPL Horizons solution of $A_2$ = (-2.121 $\pm$ 0.301)$\times10^{-13}$ au/d$^2$, and the NEODyS-2 solution of $A_2$ = (-2.241 $\pm$ 0.298)$\times10^{-13}$ au/d$^2$}. As indicated subsequently, we will compare our gravity-only solution of Kamo`oalewa with that of the JPL Horizons system.

The reliability of the Yarkovsky detection depends strongly on the orbital arc and the quality of the observations. The signal-to-noise ratio ${\rm{SNR}} = |\langle {\rm{d}}a/{\rm{d}}t \rangle/{\sigma _{\langle {\rm{d}}a/{\rm{d}}t \rangle}|} = |A_2/\sigma_{A_2}|$ is commonly used to estimate the significance \citep{farnocchia2013near}: the detections that satisfy $\rm{SNR}\ge3$ and $S\le2$ are accepted as valid detections \citep{chesley2015direct, del2018detecting}, where
\begin{equation}
    \label{eq:dadt_exp}
    S=\left| {\frac{\langle {\rm{d}}a/{\rm{d}}t \rangle}{\langle {\rm{d}}a/{\rm{d}}t \rangle_{\rm{exp}}}} \right|
\end{equation}
is an indicator parameter used to assess whether the derived drift $\langle {\rm{d}}a/{\rm{d}}t \rangle$ is physically possible and $\langle {\rm{d}}a/{\rm{d}}t \rangle_{\rm{exp}}$ is the expected value that can be estimated by scaling the parameters of Bennu \citep{del2018detecting}.

\section{Analysis of orbital fit}
\label{Sec3}
\subsection{Orbital characteristics}
\label{sec:Orbital characteristics}
To better understand the detection of the Yarkovsky effect, the orbital characteristics of Kamo`oalewa should be investigated. As mentioned by \cite{sharkey2021lunar}, the semi-major axis exhibits periodical oscillations with a longer period of $\sim$40 years due to the quasi-satellite resonance with Earth. For a better measurement of the Yarkovsky effect, a larger orbital drift over time under the influence of the Yarkovsky effect is preferred.

\begin{figure*}
    \centering
    \includegraphics[width=\textwidth]{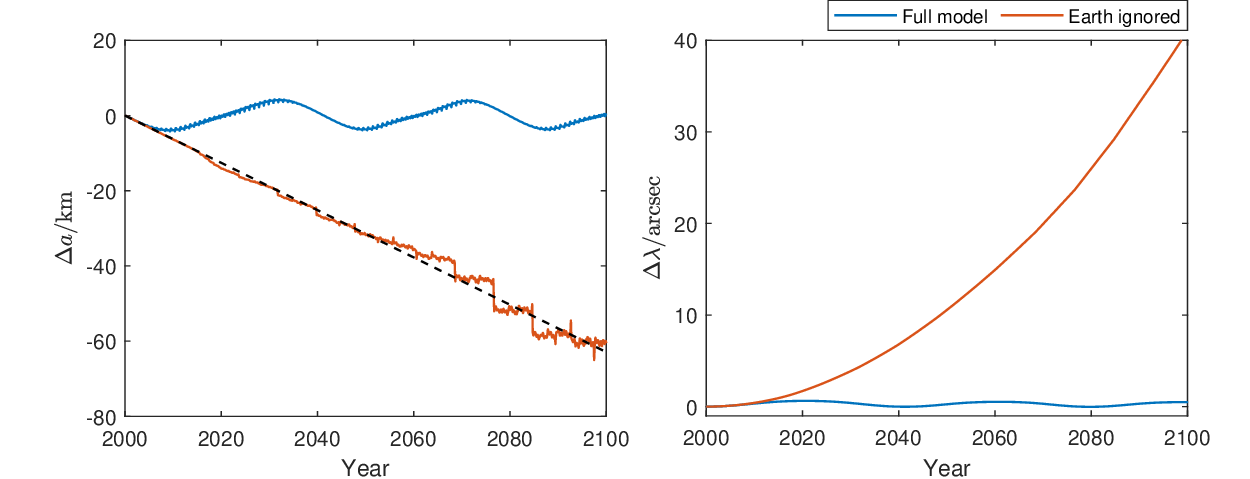}
    \caption{The two panels show the changes in $\Delta a$ and $\Delta \lambda$ for Kamo`oalewa from 2000 to 2100 with the same initial states. $\Delta a$ and $\Delta \lambda$ are the differences of the semi-major axis and the mean longitude between the Yarkovsky-included and the gravity-only models. The difference between the blue lines and the red lines is that the former includes all the planetary gravitational perturbations while the latter includes all of those but the Earth. The black dashed line is the $\Delta a$ computed with the analytical method (Eq. \ref{eq:dadt}). The value of $A_2$ in these figures is assumed to be -$1.0\times10^{-13}$ au/d$^2$.}
    \label{fig:fig00_HO3_haveYar}
\end{figure*}

The two panels in Fig. \ref{fig:fig00_HO3_haveYar} show the time evolution of the differences in the semi-major axis and the mean longitude (denoted as $\Delta a$ and $\Delta \lambda$, respectively) between the Yarkovsky-included and gravity-only models with the same initial states from 2000 to 2100 for an assumed value of $A_2=-1.0\times10^{-13}$ au/d$^2$. Two scenarios are considered in Fig.\ref{fig:fig00_HO3_haveYar}, where one includes the full planetary gravitational perturbations (as described in Section \ref{sub:dynamical model}), whereas the other model considers all of the perturbation except the gravity of Earth. Fig.\ref{fig:fig00_HO3_haveYar} exhibits that when the Earth's orbital perturbation is excluded, the drifts in the semi-major axis and the mean longitude caused by the Yarkovsky effect are shown to vary linearly and quadratically with time, respectively. Moreover, the difference of the semi-major axis $\Delta a$ is in good agreement with that predicted by the analytical method (Eq. \ref{eq:dadt}). In this situation, Kamo`oalewa would not move in the quasi-satellite orbit but behave like a normal asteroid beyond the mean-motion resonances.

However, for the real scenario considering the Earth, we can see that both the variations of $\Delta a$ and $\Delta \lambda$ are constrained and oscillate with a longer period of $\sim$40 years over time. This phenomenon can be understood that the motion of Kamo`oalewa is actually stably trapped near the Earth due to the quasi-satellite resonance, and the small perturbation of the Yarkovsky effect cannot destabilize this state. In fact, \cite{kortenkamp2013trapping} showed that the orbit of interplanetary dust particles can be trapped in the 1:1 co-orbital resonance with Earth when the influence of the non-gravitational effects including the radiation pressure, Poynting-Robertson light drag, and solar wind drag are considered. Here we show that the influence of the Yarkovsky effect can have a similar effect in the quasi-satellite regime.

\begin{figure*}
    \centering
    \includegraphics[width=\textwidth]{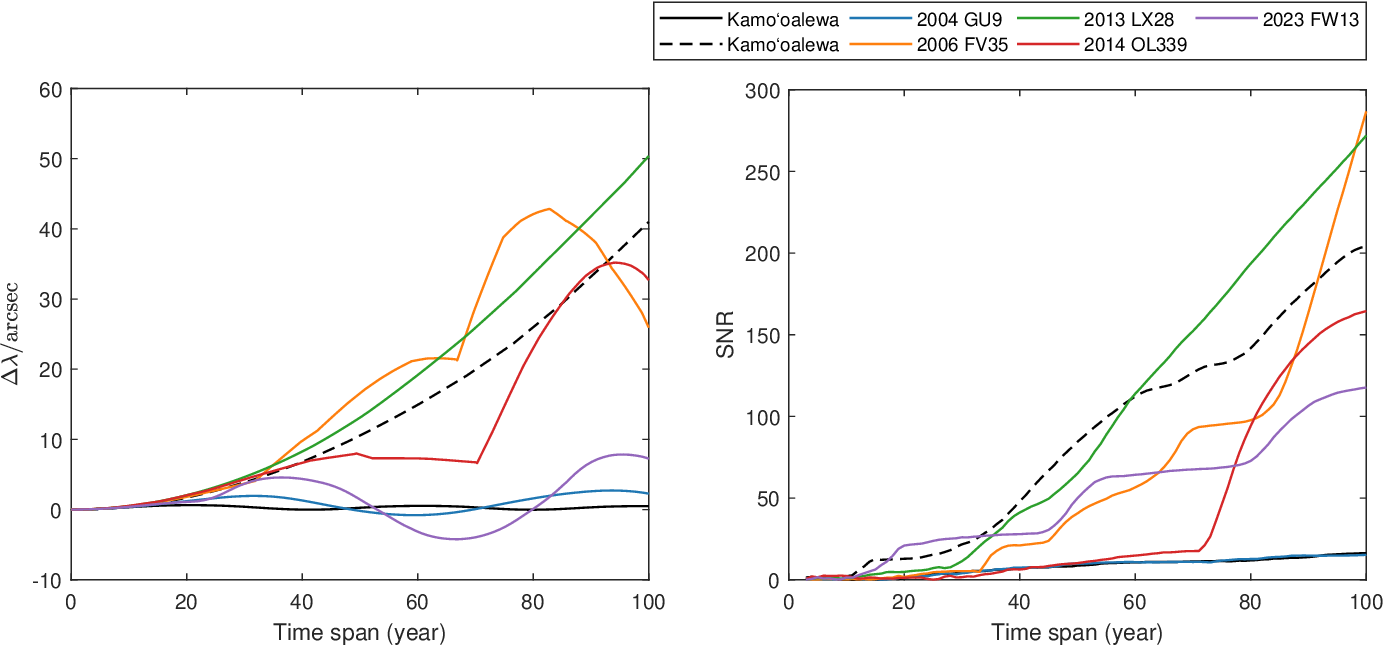}
    \caption{The left panel shows the variations of $\Delta \lambda$ for the six Earth quasi-satellites over 100 years with the same initial states. The right panel shows the evolution of the Yarkovsky detection $\rm{SNR}$ with different observational timespans with simulated optical astrometry for the quasi-satellites. The initial epochs are set to 1950 January 1 for 2013 LX28 and 2000 January 1 for other five asteroids, to ensure that the 100-year timespans are within the quasi-satellite periods. The black dashed lines denote those results for Kamo`oalewa without considering the gravity of Earth. Here $A_2$ is assumed to be -$1.0\times10^{-13}$ au/d$^2$ here.}
    \label{fig:fig03_QS_dinggui_simu}
\end{figure*}

To better understand the stability of Kamo`oalewa's orbit due to the Yarkovsky effect, a comparison of the variations of $\Delta \lambda$ over 100 years with five Earth quasi-satellites discovered so far, i.e., (164207) 2004 GU9, (277810) 2006 FV35, 2013 LX28, 2014 OL339 and 2023 FW13, is illustrated in the left panel of Fig.~\ref{fig:fig03_QS_dinggui_simu}, in which we find that only 2004 GU9 and {2023 FW13} show obvious oscillations similar to Kamo`oalewa, while the variation amplitude of Kamo`oalewa remains smallest. For the other three cases, however, they still exhibit remarkable drifts. The dynamical mechanism caused this difference will be investigated in future. This observation demonstrates that Kamo`oalewa's orbit is the most stable one in terms of the Yarkovsky perturbation among the six quasi-satellites.

It is noteworthy that this feature of Kamo`oalewa may hinder the Yarkovsky detection using ground-based astrometry, although the situation may be complicated since the detection $\rm{SNR}$ depends not only on orbital drift in the heliocentric frame, but also the observational conditions related to ground-based stations, such as the distance to the asteroid, the number of observations, the observational accuracy, etc. Nevertheless, to analyze this feature quantitatively, we carry out a series of orbital fitting including the Yarkovsky effect for the six quasi-satellites using simulated observations.

The values of $A_2$ for all the asteroids are assumed to be -$1.0\times10^{-13}$ au/d$^2$ in our simulations. In addition, the asteroids are observed from the geocentric framework, which can be a good approximation of a ground-based station due to a small parallax (typically less than 1') relative to the asteroids. The observational accuracy of the station is assumed to be 0.3", and constraints from the limit magnitude, asteroid-Sun and asteroid-Moon angles are not considered. The simulated astrometric observations are carried out on the brightest day of a year, with four observations made at 10-minute intervals for each batch of data. We fix the start observation time and vary only the end time for each case. The evolution of the Yarkovsky detection $\rm{SNR}$ over the observation time span between 3-100 years is obtained from the orbital fit, as shown in the right panel of Fig. \ref{fig:fig03_QS_dinggui_simu}, from which we can see that the $\rm{SNR}$ increases with time as expected. However, the change rate for Kamo`oalewa (as well as 2004 GU9) is much smaller than those of the other asteroids, which clearly demonstrates that the peculiar orbital characteristics of Kamo`oalewa leads to a reduced Yarkovsky detection $\rm{SNR}$.

\subsection{Results of orbital fit}
{Although we have demonstrated that the orbital drift induced by the Yarkovsky effect for Kamo`oalewa is alleviated by strong gravitational perturbation of the Earth,} it is still possible to obtain a valid Yarkovsky detection if the observations are qualified, as shown by the simulations in Fig. \ref{fig:fig03_QS_dinggui_simu}. By implementing the latest debiasing technique of \cite{eggl2020star} in the open source OrbFit 5.0.7 package {(http://adams.dm.unipi.it/orbfit/)} \citep{orbfit2011orbfit} and using the weighting scheme of \cite{verevs2017statistical}, \cite{liu2022yarkovsky} derived $A_2 = (-1.434\pm0.410)\times10^{-13}$ au/d$^2$ for Kamo`oalewa, whose $\rm{SNR}$ is greater than the threshold of 3. Here we also conduct the 7-dimensional correction with the same observations with our SBORD package and detect  $A_2 = (-1.470\pm0.400)\times10^{-13}$ au/d$^2$, with a variation of less than 3\% in the nominal value, as shown in Tab. \ref{tab:Yarkovsky-included} (where the solution is denoted as ${\rm{S}}_7^*$). {The difference between two results is marginal, which may be caused by the implementation of the force models, the propagators and the settings in the convergence schemes involved, etc.}

\begin{figure*}
    \centering
    \includegraphics[width=\textwidth]{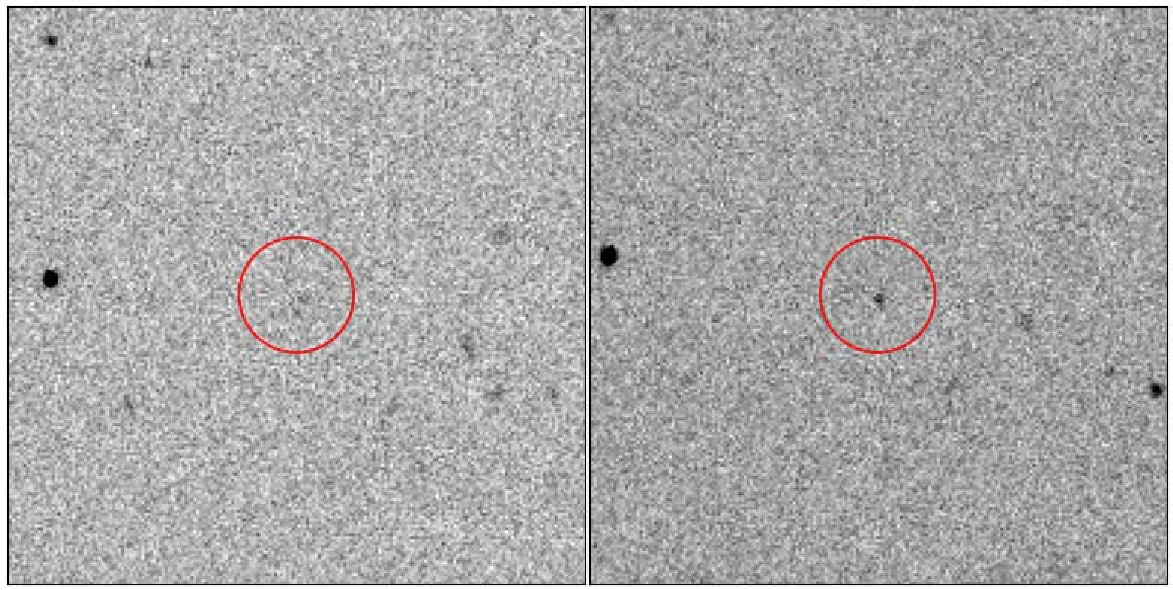}
    \caption{The two images of Kamo`oalewa (inside the {red} circles) observed by SDSS. The left was observed at 11:16:48.5 and the right at 11:21:35.1 UTC on 2004 March 17.}
    \label{fig:fig00-SDSS-fits}
\end{figure*}

{Among the observations, however, a pair of precoveries released by the Apache Point-Sloan Digital Sky Survey (SDSS) on 2004 March 17 should be carefully examined}. These data significantly extend the observational arc length from $\sim$11 to $\sim$18 years, which may give rise to an inordinate estimation on the Yarkovsky effect, as explained by \cite{chesley2015direct}. Here we can obtain the two source images at the observation epochs from the SDSS archive server\footnote{https://dr12.sdss.org/}, as shown in Fig. \ref{fig:fig00-SDSS-fits}. The inferred positions of Kamo`oalewa are circled in {red}, from which we can observe that the signal-to-noise ratio of the left image may not be sufficient for precise targeting. The reason is that the apparent magnitude of Kamo`oalewa when observed by the SDSS was $\sim$22.8, which is close to the limit of the SDSS facility \citep{raymond2004strategy}. Since the astrometric quality depends significantly on the brightness of a target \citep{verevs2017statistical}, and a moving object has to be detected from at least two valid images, we propose to reject the data pair from the orbital fit out of a conservative rule.

{Actually, we have two more points to support this treatment. First, we can perform the orbital fit using the data without the SDSS observations (the solution $\rm{S_7}$ in Tab. \ref{tab:Yarkovsky-included}, as will be discussed in the following) and propagate the orbit to compute the RA/DEC at the time of the SDSS observations. The resulting O-C is 0.23" and 0.24" for RA and -0.78" and -0.97" for DEC, respectively. Although the predictions for RA are satisfactory, it is noteworthy that the differences in DEC reach up to 2.6 and 3.2 standard deviations (the weight for the SDSS is 0.3"). Second, we need to use the star catalog flag (located in the 72nd character of the MPC 80-column observation) to estimate debiasing. However, the two SDSS observational datasets do not contain the flag information, making it impossible for us to conduct the debiasing, which increases the uncertainty of the observations. Anyhow, the rejection treatment will increase the orbital uncertainty estimate, but it is still worth considering, since having some margin in the ephemeris is usually a good idea for a space mission.}

\begin{table*}
    \centering
    \caption{Yarkovsky-included solutions and the 1-$\sigma$ uncertainties at 2023 Feb. 25 00:00:00 TDB in the J2000 ecliptic system.}
      \begin{tabular}{lccc}
      \toprule
      Para. & This work (${\rm{S}}_7^*$) & This work (${\rm{S}}_7$)  &  Unit \\
      \hline
      Semi-major axis ($a$) & 1.001007386 $\pm$ 1.5$\times10^{-8}$ & 1.001007397 $\pm$ 1.6$\times10^{-8}$ & au \\
      Eccentricity ($e$) & 0.102825645 $\pm$ 3.9$\times10^{-7}$ & 0.102825664 $\pm$ 3.9$\times10^{-7}$ & - \\
      Inclination ($i$) & 7.79201870 $\pm$ 2.7$\times10^{-5}$ & 7.79202059 $\pm$ 2.7$\times10^{-5}$ & deg \\
      Longitude of ascending node ($\Omega$) & 65.90561406 $\pm$ 2.1$\times10^{-5}$ & 65.90561102 $\pm$ 2.1$\times10^{-5}$ & deg \\
      Argument of perihelion ($\omega$) & 305.35162823 $\pm$ 2.6$\times10^{-5}$ & 305.35162629 $\pm$ 2.6$\times10^{-5}$ & deg \\
      Mean anomaly ($M$) & 141.14310070 $\pm$ 3.6$\times10^{-5}$ & 141.14308147 $\pm$ 3.7$\times10^{-5}$ & deg \\
      Position ($x$) & -150252517.8 $\pm$ 35.6 & -150252498.2 $\pm$ 37.6 & km \\
      Position ($y$) & 57277763.1 $\pm$ 23.0 & 57277816.5 $\pm$ 36.9 & km \\
      Position ($z$) & 21969115.7 $\pm$ 81.6 & 21969121.6 $\pm$ 81.8 & km \\
      Velocity ($v_x$) & -11.2749657 $\pm$ 4.6$\times10^{-6}$ & -11.2749753 $\pm$ 6.9$\times10^{-6}$ & km/s \\
      Velocity ($v_y$) & -24.9419584 $\pm$ 1.1$\times10^{-5}$ & -24.9419539 $\pm$ 1.1$\times10^{-5}$ & km/s \\
      Velocity ($v_z$) & 0.0150928 $\pm$ 8.1$\times10^{-7}$ & 0.0150941 $\pm$ 1.1$\times10^{-6}$ & km/s \\
      $A_2$ & -1.470 $\pm$ 0.400 & -1.075 $\pm$ 0.447 & 10$^{-13}$ au/d$^2$ \\
      $\langle {\rm{d}}a/{\rm{d}}t \rangle$ & -63.06 $\pm$ 17.15 & -46.10 $\pm$ 19.18 & 10$^{-4}$ au/My \\
      \hline
      \end{tabular}
      \begin{tablenotes}
        \item[1] {Note: The orbital solutions are initially determined in the middle epoch of the observation arc and then propagated to near the current epoch.}
    \end{tablenotes}
    \label{tab:Yarkovsky-included}
\end{table*}

\begin{figure*}
    \centering
    \includegraphics[width=0.85\columnwidth, height=6cm]{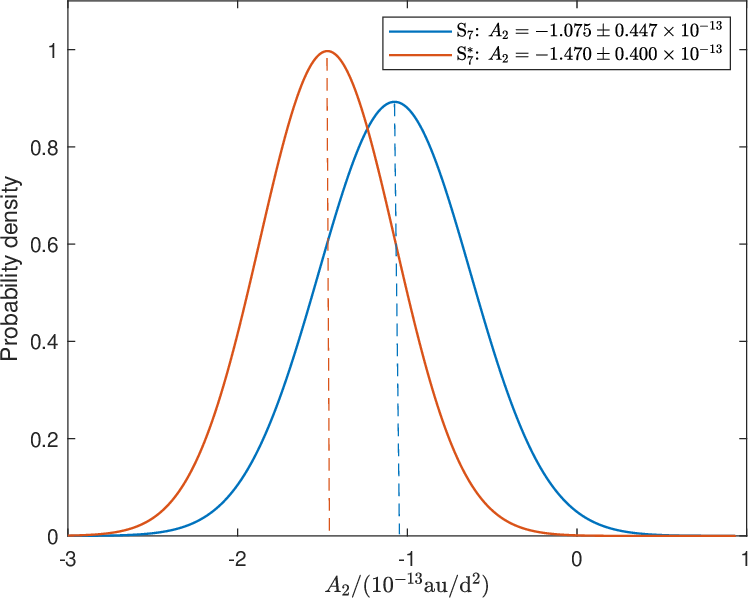}
    \caption{The probability density distributions of $A_2$ for the solutions $\rm{S_7}$ and $\rm{S_7^*}$.}
    \label{fig:fig01_Yarkovsky_bijiao}
\end{figure*}


Under the consideration above on the data quality, the solution is recomputed and denoted as $\rm{S_7}$ in Tab. \ref{tab:Yarkovsky-included}, which gives $A_2=(-1.075\pm0.447) \times10^{-13}\rm{au/d^2}$, or equivalently, $\langle {\rm{d}}a/{\rm{d}}t \rangle=(-46.10 \pm 19.18)\times10^{-4}$ au/My. The corresponding $\rm{SNR}$ is 2.4, slightly below the threshold of 2.5 for a weak detection. However, we conservatively accept this result as a weak detection, as the weights for the data are typically not that precise and a gravity-only solution may underestimate the orbital uncertainty if the Yarkovsky effect is ignored. Fig. \ref{fig:fig01_Yarkovsky_bijiao} shows the respective probability density distributions of $A_2$ for the solutions $\rm{S_7}$ and $\rm{S_7^*}$, from which we can see that $\rm{S_7}$ has a slightly lower estimate for the Yarkovsky drift than that of $\rm{S_7^*}$, as well as a slightly larger uncertainty. Nevertheless, the result still indicates that $A_2$ is likely negative, suggesting Kamo`oalewa is probably a retrograde rotator, which is consistent with that of \cite{liu2022yarkovsky}.

Note that the JPL Horizons online ephemeris system only presents gravity-only solution (with a 6-dimensional fit) for Kamo`oalewa ($\rm{S_6^{J}}$), which is listed in Tab. \ref{tab:Gravity-only}. Using the above conservative rejection scheme, we can also compute a gravity-only solution, which is displayed in Tab. \ref{tab:Gravity-only} ($\rm{S_6}$). Note that the difference in the nominal position between $\rm{S_6^{J}}$ and $\rm{S_6}$ is only $\sim$6 km, much smaller than the 1-$\sigma$ uncertainty. However, if the SDSS data are included, the solution (denoted as $\rm{S_6^*}$) has a much larger positional difference of $\sim$145 km with $\rm{S_6^{J}}$. These comparisons imply that the JPL solution {may have de-weighted the SDSS data}, which lends support to our conservative consideration on the data quality to some extent.\footnote{We note that the estimated uncertainties for the JPL solution are slightly larger than ours (the reason is unknown). However, these differences are not significant given the uncertainty in the accuracy of the observations.}.

\begin{table*}
    \centering
    \caption{Comparisons of gravity-only solutions with JPL Horizons at 2023 Feb.25 00:00:00 TDB in the J2000 ecliptic system.}
      \begin{tabular}{lcccc}
      \toprule
      Para. & JPL ($\rm{S_6^{J}}$) & This work ($\rm{S_6}$) & This work ($\rm{S_6^*}$) &  Unit \\
      \hline
      $a$ & 1.001007435 $\pm$ 3.5$\times10^{-9}$ & 1.001007435 $\pm$ 2.8$\times10^{-9}$ & 1.001007441 $\pm$ 2.1$\times10^{-9}$ & au \\
      $e$ & 0.102826447 $\pm$ 2.4$\times10^{-7}$ & 0.102826443 $\pm$ 2.1$\times10^{-7}$ & 0.102826990 $\pm$ 1.2$\times10^{-7}$ & - \\
      $i$ & 7.79207609 $\pm$ 1.7$\times10^{-5}$ & 7.79207538 $\pm$ 1.4$\times10^{-5}$ & 7.79211288 $\pm$ 8.3$\times10^{-6}$ & deg \\
      $\Omega$ & 65.90557126 $\pm$ 1.5$\times10^{-5}$ & 65.90557112 $\pm$ 1.3$\times10^{-5}$ & 65.90554701 $\pm$ 1.0$\times10^{-5}$ & deg \\
      $\omega$ & 305.35166689 $\pm$ 2.3$\times10^{-5}$ & 305.35166430 $\pm$ 2.1$\times10^{-5}$ & 305.35169827 $\pm$ 1.8$\times10^{-5}$ & deg \\
      $M$ & 141.14300832 $\pm$ 2.3$\times10^{-5}$ & 141.14300950 $\pm$ 2.1$\times10^{-5}$ & 141.14299347 $\pm$ 2.1$\times10^{-5}$ & deg \\
      $x$ & -150252552.0 $\pm$ 35.2 & -150252550.1 $\pm$ 30.8 & -150252628.9 $\pm$ 18.7 & km \\
      $y$ & 57277879.8 $\pm$ 26.2 & 57277885.1 $\pm$ 23.2 & 57277827.2 $\pm$ 14.7 & km \\
      $z$ & 21969289.4 $\pm$ 51.1 & 21969287.4 $\pm$ 43.3 & 21969400.6 $\pm$ 25.0 & km \\
      $v_x$ & -11.2749878 $\pm$ 4.5$\times10^{-6}$ & -11.2749886 $\pm$ 4.0$\times10^{-6}$ & -11.2749792 $\pm$ 2.8$\times10^{-6}$ & km/s \\
      $v_y$ & -24.9419298 $\pm$ 4.5$\times10^{-6}$ & -24.9419295 $\pm$ 3.7$\times10^{-6}$ & -24.9419201 $\pm$ 2.2$\times10^{-6}$ & km/s \\
      $v_z$ & 0.0150945 $\pm$ 1.2$\times10^{-6}$ & 0.0150946 $\pm$ 1.0$\times10^{-6}$ & 0.0150925 $\pm$ 8.0$\times10^{-7}$ & km/s \\
      \hline
      \end{tabular}
    \label{tab:Gravity-only}
\end{table*}

\begin{figure*}
    \centering
    \includegraphics[width=\textwidth]{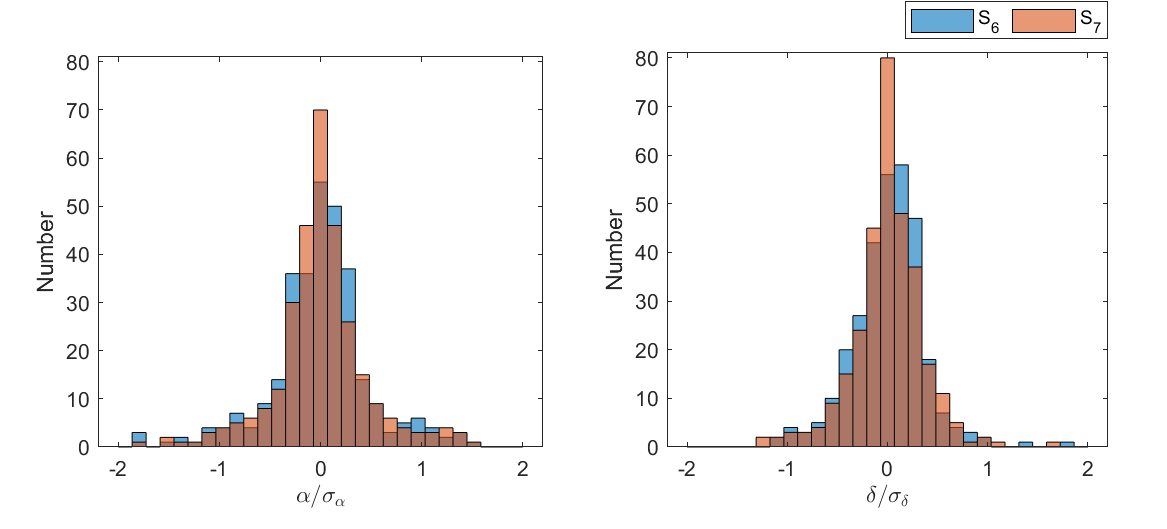}
    \caption{Histograms of normalized O-C residuals of RA ($\alpha$) and DEC ($\delta$) for the solution $\rm{S}_6$ and $\rm{S}_7$.}
    \label{fig:fig10_HO3_dinggui_cancha}
\end{figure*}

Fig. \ref{fig:fig10_HO3_dinggui_cancha} presents the histograms of normalized O-C residuals for the solution $\rm{S}_6$ and $\rm{S}_7$, from which we note that the distributions of the residuals are more concentrated around the zero point when the Yarkovsky effect is introduced. In fact, the normalized ${\chi ^2}$ is 0.38752 and 0.36985 for $\rm{S}_6$ and $\rm{S}_7$, respectively. Therefore, the latter is a better fit. In the following discussions we will analyze the orbital uncertainty based on the solution ${\rm{S}}_7$.

\section{Analysis of orbital uncertainty}
\label{Sec4}
\subsection{Orbital uncertainty in different frames}
The orbital uncertainty at the initial epoch $t_0$ can be obtained from the covariance matrix $P\left( {{t_0}} \right)$, which is solved together with the orbit determination procedure. According to the linear covariance propagation theory, the covariance matrix $P(t)$ at time $t$ is given by
\begin{equation}
    \label{eq:covariance}
    \left\{ \begin{array}{l}
        P\left( t \right) = \Phi \left( {t,{t_0}} \right)P\left( {{t_0}} \right)\Phi {\left( {t,{t_0}} \right)^T}\\
        \Phi \left( {t,{t_0}} \right) = \frac{{\partial X\left( t \right)}}{{\partial X\left( {{t_0}} \right)}}
        \end{array} \right.
\end{equation}
in which $X$ is the 7-dimensional vector $X=[Z, {A_2}]$ ($Z$ represents the orbital parameters, which can be the Cartesian positions and velocities or the orbital elements). $\Phi \left( {t,{t_0}} \right)$ is the 7$\times$7 state transition matrix (STM) that is derived numerically by integrating the variational equations \citep{montenbruck2002satellite, milani2010theory}. If the Cartesian coordinates are used, we can transform the covariance matrix $P(t)$ to another reference system by
\begin{equation}
    P'\left( t \right) = B \cdot P\left( t \right){B^T}
\end{equation}
in which $P'\left( t \right)$ is the covariance in the new reference, and $B$ is the rotation matrix between the two references. Note that the covariance remains unchanged if we apply a translation to an {inertial} frame.

\begin{figure*}
    \centering
    \includegraphics[width=\textwidth]{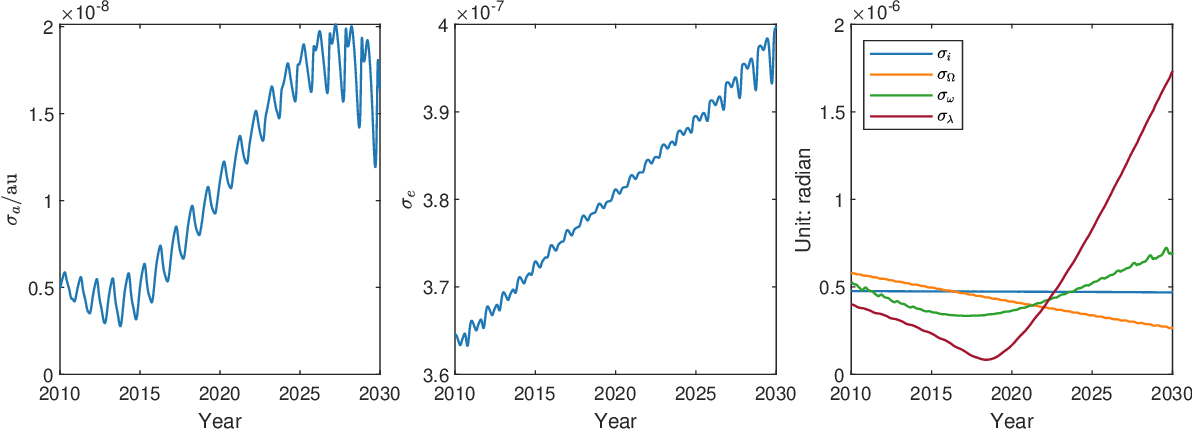}
    \caption{Time evolution of uncertainty in the orbital elements from 2010 to 2030 for the solution $\rm{S_7}$.}
    \label{fig:fig04_rms_Z}
\end{figure*}

Fig. \ref{fig:fig04_rms_Z} shows the time evolution of the uncertainty in the orbital elements from 2010 to 2030 for our solution $\rm{S_7}$, from which we can see that $\sigma_a$, $\sigma_e$ and $\sigma_i$ are relatively smaller compared to the results of the other angular elements $\sigma_{\Omega}$, $\sigma_{\omega}$ and $\sigma_{\lambda}$, and $\sigma_{\lambda}$ is generally the maximum one among the six. We find that $\sigma_{\lambda}$ reaches a minimum value in $\sim$2018, and then gradually increases superlinearly with time, which is attributed to the uncertainty in the Yarkovsky effect.

\begin{figure*}
    \centering
    \includegraphics[width=0.85\textwidth]{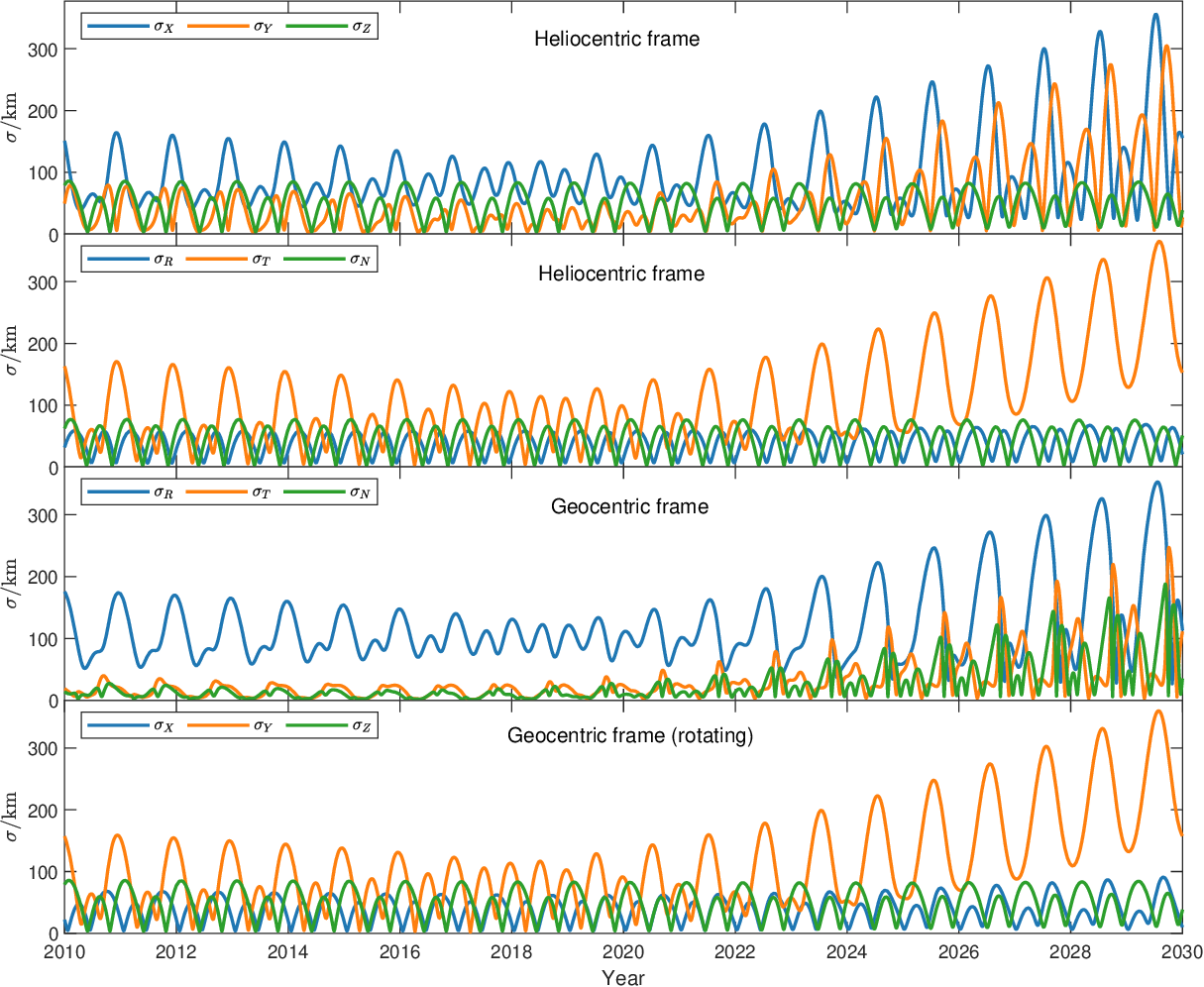}
    \caption{Variations of the positional uncertainty from 2010 to 2030 for the solution $\rm{S_7}$ in different frames. The two top panels show the changes in the heliocentric frame for the Cartesian and RTN coordinates, respectively. The third panel shows the results for the RTN directions in the geocentric frame and the bottom panel shows the results for the Cartesian coordinates in the geocentric Sun-Earth rotating frame.}
    \label{fig:fig04_rms_XYZ_Yark_diff_reference}
\end{figure*}

In addition, Fig. \ref{fig:fig04_rms_XYZ_Yark_diff_reference} shows the time evolution of the positional uncertainty (1-$\sigma$) for Kamo`oalewa. The results in the four panels are the uncertainty of the Cartesian coordinates in the heliocentric frame, the RTN uncertainty in the heliocentric frame, the RTN uncertainty in the geocentric frame, and the Cartesian uncertainty in the geocentric rotating frame, respectively. We can see that all the results exhibit complicated cyclic changes with time, except that the uncertainty from 2010 to 2020 is generally smaller than that from 2020 to 2030, which can be explained by the fact that the orbit is constrained by the astrometry from 2010 to 2019 (the constraint by the observations in 2021 is weak due to a relatively lower accuracy estimation).

{According to the results in the two middle panels, we can see that the uncertainty from 2010 to 2020 is mainly concentrated along the radial direction of the orbit in the geocentric frame. This is because ground-based optical astrometry can provide a better constraint on the orbit along the geocentric transverse and normal directions, while the constraint for the radial direction is weaker.} However, the uncertainty from 2020 to 2030 is more suitable for description in the heliocentric frame, where the transverse uncertainty gradually increases to a maximum of $\sim$370 km in 2030 while the radial and normal components remain within $\sim$80 km.

\begin{figure}
    \centering
    \includegraphics[width=0.47\textwidth]{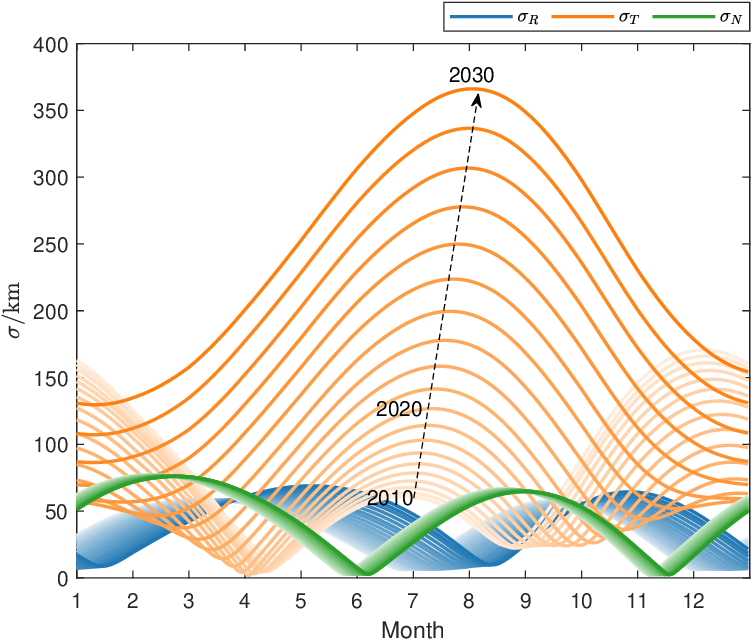}
    \caption{Plots of the positional uncertainties for the solution $\rm{S_7}$ from 2010 to 2030 in the heliocentric RTN directions, in which the horizontal axis is taken as the month, and the color gradient from dark to light indicates the evolution of the year. The dashed line arrow shows the evolution direction for $\sigma_T$.}
    \label{fig:fig06_rms_STW_season}
\end{figure}

Moreover, monthly dependencies in the uncertainty can be observed from the oscillations in Fig. \ref{fig:fig04_rms_XYZ_Yark_diff_reference}. To better illustrate this phenomenon, Fig. \ref{fig:fig06_rms_STW_season} shows the evolution in the heliocentric RTN directions, with the month as the horizontal axis. The results clearly demonstrate the oscillations of the uncertainties within a year. However, due to the complicated orbital evolution relative to the Earth, the month at the maximum amplitude shifts slowly with the year. For the period 2025-2027, which is more important for the mission, the uncertainty reaches a minimum in around January and a maximum in around July. As seen in the upper right panel of Fig. \ref{fig:fig00_HO3_orbit_when_obs}, the orbit of Kamo`oalewa moves approximately towards the leading and trailing edges in January and July, respectively. The normal component always reaches a minimum around late May (or early June) or mid-November, which occur when the orbit is moving around the ascending/descending nodes, or crossing the ecliptic.

\subsection{Uncertainty propagation in the quasi-satellite regime}
The previous subsection only analyzes the uncertainty propagation for a few years. However, due to the peculiar orbital configuration of Kamo`oalewa, it will be helpful to improve the knowledge of the orbit propagation by investigating the uncertainty in the quasi-satellite regime for a longer period of time.

\begin{figure*}
    \centering
    \includegraphics[width=0.90\textwidth]{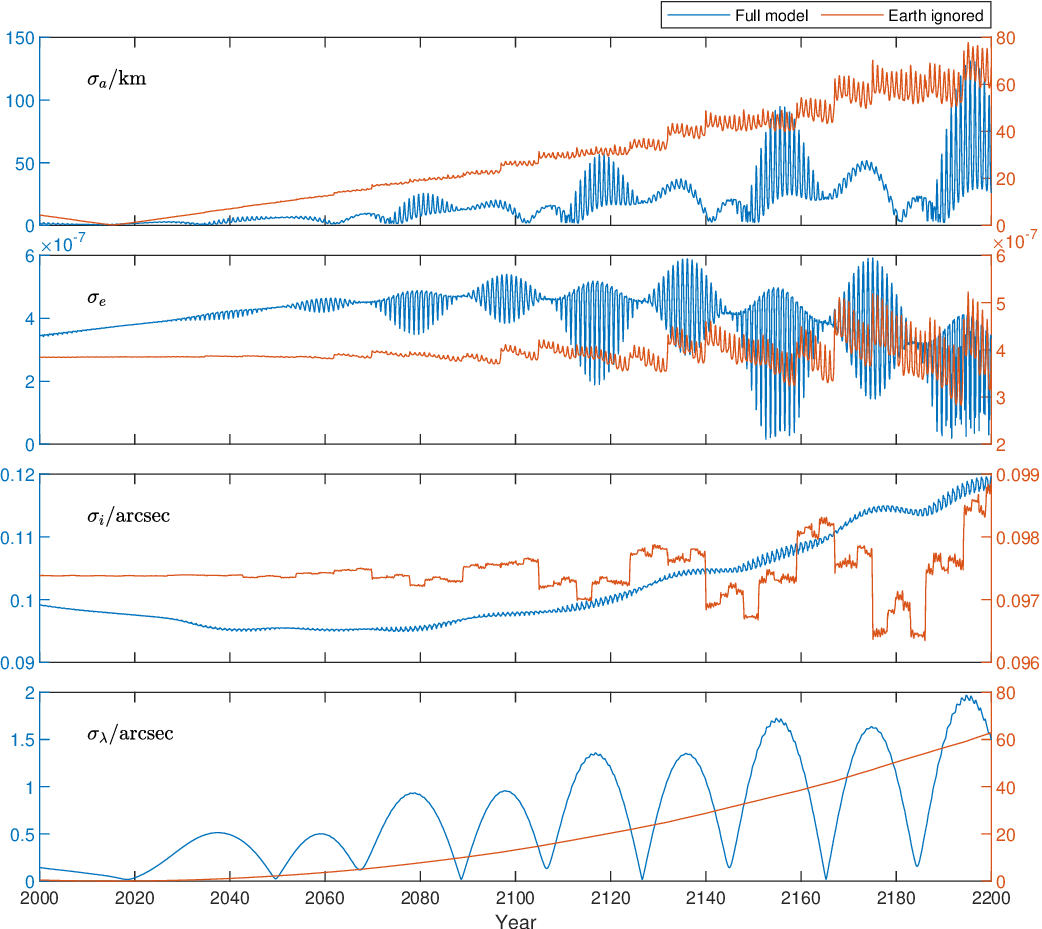}
    \caption{The panels show the uncertainty in the semi-major axis, eccentricity, inclination, and the mean longitude, respectively, from 2000 to 2200. The results in blue are computed using the full planetary gravitational perturbation model and the red ones excludes the Earth perturbation.}
    \label{fig:fig08_rmsZ_Ear_noEar_Yar}
\end{figure*}

From Eq. \ref{eq:covariance}, the covariance matrix depends strongly on the STM, which is solved together with the orbit propagation. Similar to the study in the Section \ref{sub:method of yarkovsky effect detection}, here we consider two force models to compute the STM in addition to the Yarkovsky force, where one takes into account the full planetary gravitational perturbations, and the other includes those except the Earth. Fig. \ref{fig:fig08_rmsZ_Ear_noEar_Yar} shows the uncertainties of the semi-major axis, eccentricity, inclination, and mean longitude, respectively, for the two models, in which the initial covariances are taken from the solution $\rm{S_7}$. The time range is from 2000 to 2200, well within the quasi-satellite period.

The results demonstrate that the along-track uncertainty $\sigma_{\lambda}$ is undoubtedly the most significant component for both the models. However, the evolution differs significantly between them: for the full model, $\sigma_{\lambda}$ undergoes complex oscillatory changes with increasing amplitudes over time, while for the model without Earth the values increase quadratically at much faster rates. The difference in $\sigma_{\lambda}$ between the both can be as large as 40 times in 2200.

\begin{figure*}
    \centering
    \includegraphics[width=\textwidth]{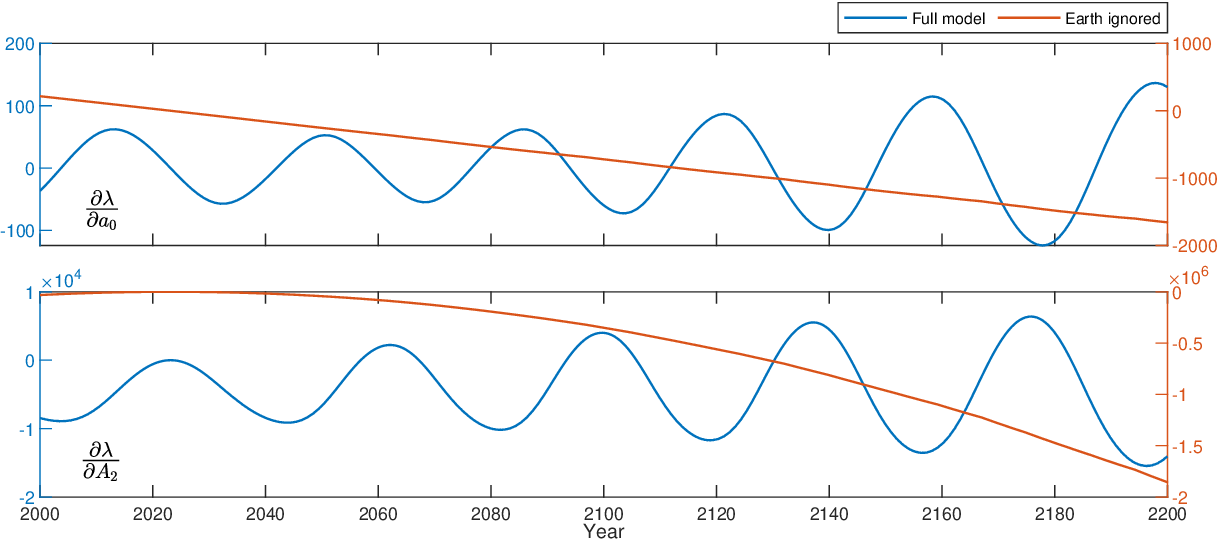}
    \caption{The time evolution of $\frac{{\partial \lambda \left( t \right)}}{{\partial {a_0}}}$ and $\frac{{\partial \lambda \left( t \right)}}{{\partial {A_2}}}$ for the two force models from 2000 to 2200, in which the initial epoch is 2023 Feb. 25 00:00:00 TDB.}
    \label{fig:fig09_propagation_ztzy}
\end{figure*}

The propagation of $\sigma_{\lambda}$ is mainly governed by the terms $\frac{{\partial \lambda \left( t \right)}}{{\partial {a_0}}}$ and $\frac{{\partial \lambda \left( t \right)}}{{\partial {A_2}}}$ in the STM, which correspond to the uncertainties in the initial semi-major axis and the Yarkovsky effect, respectively. Fig. \ref{fig:fig09_propagation_ztzy} shows the time evolution of the two partial derivatives for the two models. We find that the magnitudes of the variations in $\frac{{\partial \lambda \left( t \right)}}{{\partial {A_2}}}$ are significantly larger than that of $\frac{{\partial \lambda \left( t \right)}}{{\partial {a_0}}}$ for both the models, indicating that the uncertainty caused by the Yarkovsky effect is much larger than that of the uncertainty in the initial semi-major axis for this propagation of 200 years. Therefore, we will focus on the evolution of $\frac{{\partial \lambda \left( t \right)}}{{\partial {A_2}}}$ in the following.

As expected, for the model without Earth, $\frac{{\partial \lambda \left( t \right)}}{{\partial {A_2}}}$ propagates quadratically with time. For the full model, however, the evolution oscillates over a period of about 40 years, with the amplitude increasing slowly with time. This very different mode of change explains the remarkable difference in the evolution of $\sigma_{\lambda}$. Note that Kamo`oalewa would not behave as a quasi-satellite when the perturbation of Earth is excluded. This phenomenon indicates that the propagation of the orbital uncertainty under the Yarkovsky effect is strongly constrained, which facilitates the orbital prediction throughout the quasi-satellite period.

To understand the oscillatory variation in $\sigma_{\lambda}$ for the full model, we use the following forward difference method to approximately calculate the partial derivative
\begin{equation}
    \label{eq:dlambda_dh}
\frac{{\partial \lambda \left( t \right)}}{{\partial {A_2}}} \approx \frac{{\lambda \left( {{Z},{A_2} + h} \right) - \lambda \left( {{Z},{A_2}} \right)}}{h}
\end{equation}
in which $h$ is a small quantity. Since $A_2$ is also very small, we can safely substitute $h$ with $-A_2$ and then we have
\begin{equation}
    \label{eq:dlambda_dh2}
    \frac{{\partial \lambda \left( t \right)}}{{\partial {A_2}}} \approx \frac{{\lambda \left( {{Z},0} \right) - \lambda \left( {{Z},{A_2}} \right)}}{{ - {A_2}}} = \frac{{\Delta \lambda }}{{{A_2}}}
\end{equation}
in which ${\Delta \lambda }$ is the difference in the mean longitude between the Yarkovsky-included and gravity-only model, a typical example of which can be seen in Fig. \ref{fig:fig00_HO3_haveYar}. This feature indicates that the evolution of the along-track uncertainty $\sigma_{\lambda}$ under the influence of the Yarkovsky effect is consistent with the evolution of ${\Delta \lambda }$, which is governed by the quasi-satellite resonance with Earth, as explained in Section \ref{sec:Orbital characteristics}.

\section{Summary and Conclusions}
\label{Sec5}
469219 Kamo`oalewa is a small near-Earth asteroid, which is currently known as the most stable quasi-satellite of Earth. In addition, this asteroid is the sample-return target of Chinese Tianwen-2 mission scheduled for launch in 2025. To facilitate the mission, precise orbit determination is important to ensure a successful rendezvous with the object. However, the small size of Kamo`oalewa may correspond to a remarkable Yarkovsky acceleration. This is usually the most important non-gravitational force that can affect the orbital precision and may be detected from the orbital fit. Considering that additional observations may not be available prior to the launch {(the apparent magnitude of the observation window at the brightest time is dimming in the next few years)}, the uncertainty estimate for the next few years including the Yarkovsky effect is needed. In addition, due to the peculiar quasi-satellite orbit, the uncertainty characteristics in the quasi-satellite regime are analyzed, which provides more insights into the orbit propagation of Kamo`oalewa.

In this work, we first show numerically that the Yarkovsky effect can cause a periodic change in Kamo`oalewa's semi-major axis due to its particular quasi-satellite resonance with Earth, rather than a normal secular drift for a general asteroid. This feature can reduce the sensitivity of the orbital drift to the influence of the Yarkovsky effect, and to some extent making it more difficult to detect the drift from the orbital fitting, as demonstrated by the comparisons of $\rm{SNR}$ over observational arc among the six Earth quasi-satellites discovered so far (Fig. \ref{fig:fig03_QS_dinggui_simu}).

Nevertheless, we still find a weak detection of the Yarkovsky effect using the ground-based optical observations spanning from {2011 to 2021 (the precoveries by SDSS in 2004 are rejected due to the poor quality)}, giving $A_2 = -1.075\pm0.447\times 10^{-13} \rm{au/d}^2$, with a slightly lower significance in the detection $\rm{SNR}$ than that of \cite{liu2022yarkovsky}.  Here we take a relatively conservative approach and reject a pair of pre-recovery SDSS observations to avoid overestimating the drift, which is supported by the possible lower signal-to-noise ratio in one of the {source images.}

Based on our derived solution including the Yarkovsky effect, the orbital uncertainty propagation of Kamo`oalewa is extensively studied. We find that the positional uncertainty comes mainly from the radial direction for the geocentric orbit in 2010-2020 and is concentrated in the transverse direction for the heliocentric orbit in 2020-2030. The variation of the uncertainty within a year is clearly related to the month due to the peculiar orbit configuration. The heliocentric transverse uncertainty is minimized around January and maximized around July in 2025-2027, corresponding to the period when the orbit moves toward the leading and trailing edges, respectively.

Finally, we further investigate a long-term uncertainty propagation in the quasi-satellite regime. We show that the along-track uncertainty with the Yarkovsky effect can be significantly constrained during this period, rather than a general rapid quadratic propagation. This feature indicates that the orbit precision of this asteroid is relatively stable even when the uncertainty of the Yarkovsky effect is considered.

Note that the above conclusions about the orbital uncertainty in the next few years are based on the ground-based optical observations from 2011 to 2021. Since the T12 station accounts for most of the observations, the treatment of the weighting will lead to a various solution. In particular, since the uncertainties have been inflated by a $\sqrt{N}$ factor to mitigate possible effects of unresolved systematics \citep{verevs2017statistical}, this artificial treatment will also add more uncertainties in the weight and then finally affect the solution. {With the addition of spacecraft-based radio and optical observations during the rendezvous, the uncertainty in the Yarkovsky detection and orbital precision is likely to be significantly improved.}

Nevertheless, the uncertainty analysis procedures proposed here remain valid. Moreover, since the behavior of orbital variations under the influence of the Yarkovsky effect is an intrinsic orbital property of Kamo`oalewa, the uncertainty propagation for a longer period of time in the quasi-satellite regime is still constrained, even the orbital solution would be improved in the future. This feature may also apply to other quasi-satellites (e.g., 2004 GU9 and 2023 FW13), which will be explored in forthcoming study.

\begin{acknowledgments}
We thank the referee for constructive comments and suggestions to improve the manuscript. This work is financially supported by the B-type Strategic Priority Program of the Chinese Academy of Sciences (Grant No. XDB41000000), the National Natural Science Foundation of China (Grant Nos. 11873098, 12150009, 12033010, 62227901, 12073084), the Natural Science Fundation for Young Scientists of Jiangsu Province (Grant No. BK20221162), the Space Debris and Near-Earth Asteroid Defense Research Project (Grant Nos. KJSP2020020204, KJSP2020020102) and Foundation of Minor Planets of the Purple Mountain Observatory. We appreciate Yi Xie and Steven R. Chesley for their insightful discussions.
\end{acknowledgments}

\bibliography{ms}{}
\bibliographystyle{aasjournal}

\end{document}